# A Differentiable Physical Framework for Goal-Driven Spin-State Engineering in Magnetic Resonance Spectroscopy

Gaocheng Fu[1,#], Shiji Zhang[1,#], Kai Huang[1], Xue Yang[1], Huilin Zhang[1], Daxiu Wei[1]*, Ye-Feng Yao[1,2]*

1. Physics Department & Shanghai Key Laboratory of Magnetic Resonance, School of Physics and Electronic Science, East China Normal University, North Zhongshan Road 3663, Shanghai 200062, P. R. China.
2. Institute of Medical Magnetic Resonance and Molecular Imaging Technology, East China Normal University, 200241, Shanghai, P. R. China.

*: The corresponding authors:
Daxiu Wei: dxwei@phy.ecnu.edu.cn
Ye-Feng Yao: yfyao@phy.ecnu.edu.cn
[#]: Gaocheng Fu and Shiji Zhang contributed equally to this work.

**Abstract**

Magnetic Resonance Spectroscopy (MRS) offers a unique non-invasive window into metabolic processes, yet its potential remains strictly constrained by severe spectral congestion and intrinsic insensitivity. Traditional pulse sequence design, tethered to human intuition, predominantly targets simple quantum states, thereby overlooking the vast majority of the exponentially scaling operator space which consists of complex spin superpositions. Here, we introduce a spectrum-driven, end-to-end differentiable physical framework that transcends these heuristic limitations. By integrating physical laws with automatic differentiation algorithm, our approach directly navigates the high-dimensional spin dynamics space, bypassing the intractable inverse problem of state preparation. This enables the discovery of non-intuitive, complex mixed states that simultaneously satisfy the dual objectives of selective excitation and interferometric signal enhancement. We validate this paradigm by achieving the robust separation of Glutamate and Glutamine, which is a longstanding neuroimaging challenge, in the human brain at 3T, demonstrating spectral fidelity superior to conventional methods. By unlocking the "dark" informational content of nuclear spin ensembles, our work establishes a generalizable paradigm for goal-driven quantum state engineering in magnetic resonance and beyond.

## Introduction

Magnetic Resonance Spectroscopy (MRS) stands as a powerful, non-invasive window into the biochemical processes of living systems[1]. Despite its unique capability to quantify metabolites in vivo, the full potential of MRS has long been constrained by two fundamental bottlenecks: severe spectral congestion and intrinsic low sensitivity[2]. In complex biological environments, signals from low-concentration metabolites are often buried beneath massive background noise or obscured by the overlapping resonances of abundant species. For decades, overcoming these limitations has been the central pursuit of pulse sequence engineering[3-5].

However, traditional approaches to pulse design are inherently limited by human intuition. Researchers typically restrict their focus to a narrow subset of "simple" quantum states[6-8] (e.g., single or low-order coherence operators) that are conceptually easy to visualize and manipulate. This heuristic approach leaves the vast majority of the spin dynamics space unexplored. In a full coupled spin system, the operator space scales exponentially, for a 5-spin system, this results in 1024 operators. Consequently, complex nuclear spin superpositions, which may hold the key to high selectivity and signal enhancement, remain a "cognitive blind spot," submerged in a "dark" cloud of inaccessible quantum states. Furthermore, the lack of a general mathematical solution for the inverse problem (deducing RF parameters from a target state)[9, 10] often leads to a dead end where theoretically desirable states cannot be experimentally prepared.

To bridge this gap, we introduce a spectrum-driven, differentiable physical framework for goal-oriented quantum state engineering. Departing from the conventional paradigm of targeting specific intermediate quantum states, our approach integrates the Liouville-von Neumann equation with deep learning optimization to establish an end-to-end differentiable link between RF pulse parameters and the final spectrum. By formulating the pulse design as a direct spectral loss optimization problem, we bypass the need for human intuition and allow the algorithm to freely navigate the high-dimensional spin dynamics space. This enables the discovery of non-intuitive, complex spin superpositions that simultaneously achieve selective excitation (to resolve overlap) and interferometric enhancement (to boost weak signals).

We validate this paradigm by addressing the classic challenge of distinguishing Glutamate (Glu) and Glutamine (Gln) in the human brain at 3T[11-16]. Our results demonstrate that machine-designed sequences, utilizing complex mixed states previously deemed "too complex" or "invisible," significantly outperform human-designed counterparts in both selectivity and sensitivity. This work not only provides a generalizable solution for spectral editing in MRS but also represents a fundamental shift in how we perceive and control quantum systems. By unlocking the "dark" information content within complex spin ensembles, our framework opens new avenues for advancements in magnetic resonance, quantum control, and beyond.

## Results

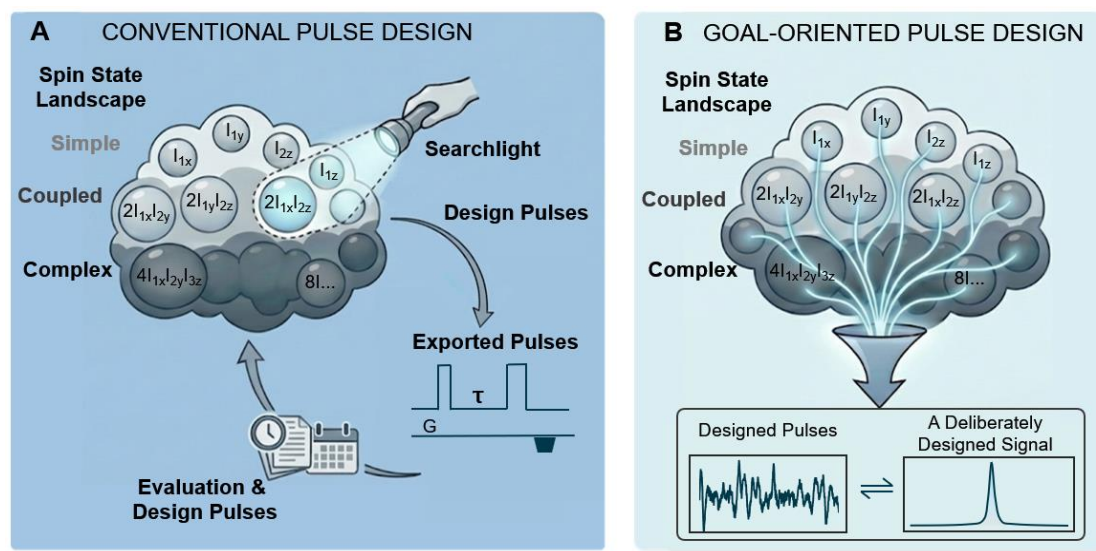

**Figure 1.** Paradigm Shift in NMR Pulse Design: From Heuristic Exploration to Goal-Driven Discovery. A) Conventional Pulse Design: Traditionally, pulse sequence engineering has been confined to a "searchlight" approach, primarily targeting simple, intuitive spin states $I_{1x}$, $I_{2z}$ within a vast landscape. Consequently, a wealth of complex spin-state manifolds, characterized by high-order correlations and rich information density, remains obscured ("submerged in the dark clouds") and underutilized. B) Goal-Driven Pulse Design via AI: In contrast, our AI-driven framework acts as a precision filter, Navigating the intricate spin-state landscape to identify and harness specific complex superpositions tailored to experimental objectives. By effectively "mining" these previously inaccessible states, the system enables the deliberate design of signals with bespoke characteristics, transforming the latent complexity of spin dynamics into a programmable resource for advanced spectroscopy.

**A New Paradigm for Pulse Design.** Traditionally, pulse sequence design has been confined to a "searchlight" approach, primarily targeting simple, intuitive spin states $I_{1x}$, $I_{2z}$ within a vast landscape. Consequently, a wealth of complex spin-state manifolds, characterized by high-order correlations and rich information density, remains obscured ("submerged in the dark clouds") and underutilized. To transcend the inherent limitations of conventional pulse sequence design in complex spin systems, we introduce a spectrum-driven, differentiable design paradigm (Fig. 1). Traditional strategies, tethered to heuristic intuition, typically restrict the search for optimal spin states to a narrow, low-order subspace that is conceptually easy to visualize (Fig. 1A). This artificial dimensionality reduction fails to capitalize on the rich information density available within the full Hilbert space. Furthermore, because deducing control field parameters from a pre-defined target state constitutes a mathematically ill-posed inverse problem, these methods often suffer from an inefficient "design-fail-refine" cycle, where theoretically desirable states prove physically inaccessible. In contrast, our approach effectuates a fundamental dimensional transformation (Fig. 1B). Rather than performing an explicit search within the exponentially scaling $4^n$-dimensional state space, we project the optimization problem onto the intuitive 2D spectral plane. By establishing an end-to-end differentiable mapping from RF parameters to the final spectral observable, the framework enables direct gradient-based optimization of the spectral profile. This process not only implicitly locates complex, non-intuitive coherence superpositions that yield optimal responses but, crucially, guarantees their physical realizability. This "what-you-see-is-what-you-get" closed-loop mode synchronizes pulse generation with spectral targeting, effectively bypassing the intractable inverse problem of state preparation and unlocking the latent potential of "dark" spin manifolds.

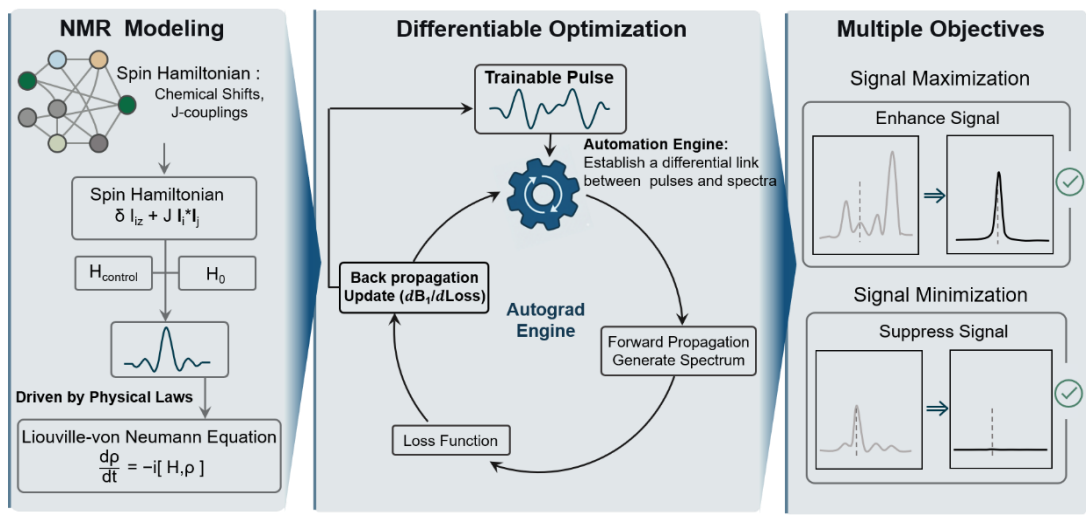

Figure 2. **Integration of Physics-Driven Modeling and Differentiable Optimization for Signal Engineering. Left: NMR Modeling.** The framework begins by encoding the spin Hamiltonian, including chemical shifts and J-couplings, into the Liouville-von Neumann equation. This ensures the optimization process is strictly governed by fundamental physical laws, specifically the evolution of the density matrix ρ over time. **Center: Differentiable Optimization.** By using the PyTorch Autograd engine[17], we establish a differential link between the pulse parameters and the resulting spectra. Through a closed-loop cycle of forward propagation (spectrum generation) and backpropagation (updating pulse), the Automation Engine iteratively refines the "Trainable Pulse" to minimize the defined loss function. **Right: Multiple Objectives.** This goal-driven design enables the sophisticated manipulation of complex spin-state manifolds. By precisely editing pulse sequences, we can achieve diverse experimental goals, such as maximizing specific resonances (Signal Enhancement) or selectively eliminating unwanted backgrounds (Signal Suppression).

To operationalize this spectrum-driven philosophy, we constructed a unified differentiable optimization framework (Fig. 2) centered on the direct embedding of the Liouville-von Neumann equation[18]. This ensures that the simulation of the density matrix's temporal evolution, governed by internal Hamiltonian interactions (chemical shifts and scalar couplings) and external RF controls, strictly adheres to fundamental quantum mechanical principles.

Building on this physical foundation, we leveraged automatic differentiation to transform the entire forward process, from spin dynamics evolution to Fourier transform sampling, into a fully differentiable computational chain[19]. This establishes a direct gradient backpropagation pathway from the final spectral observables back to the RF pulse parameters. Furthermore, the framework incorporates a flexible multi-task loss function, allowing for the direct imposition of maximization or minimization constraints on specific metabolite signals within the spectral domain. This capability enables the generation of bespoke pulse sequences that precisely satisfy sophisticated spectral editing requirements, bridging the gap between theoretical optimization and experimental utility.

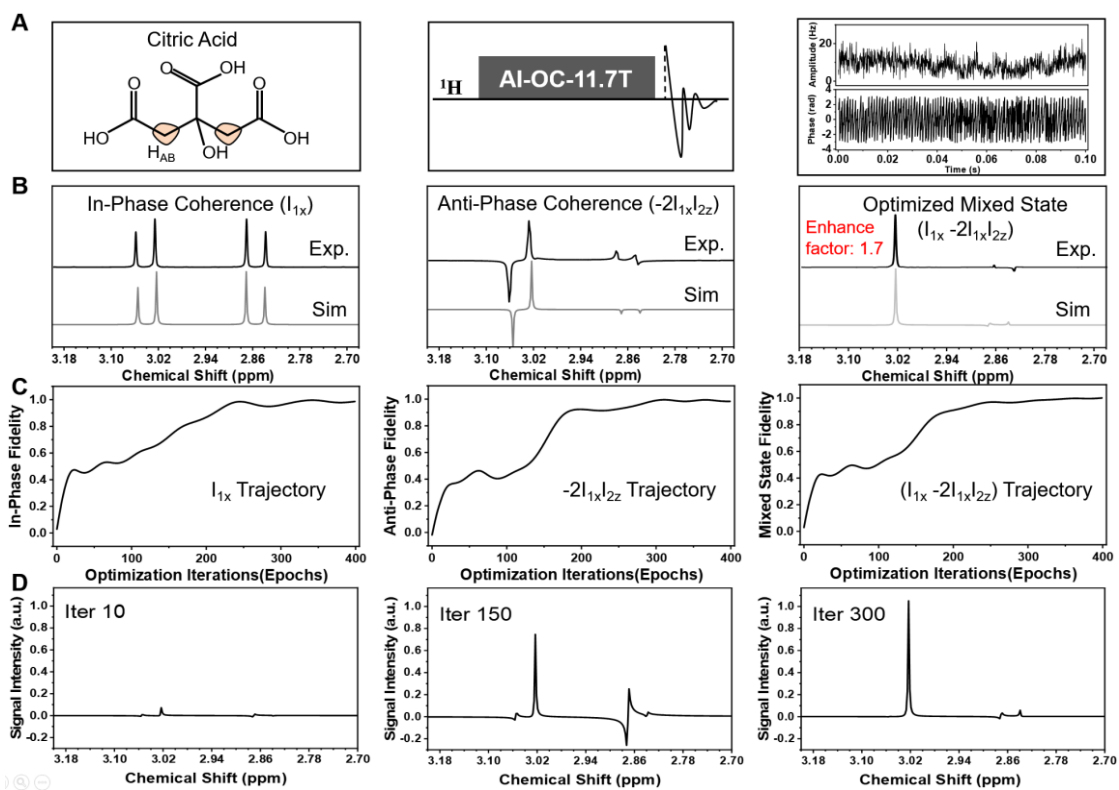

Figure 3. **Experimental Validation of Goal-Driven Pulse Design on Citric Acid**. **A) Model Molecule and Pulse Architecture. Left:** The study utilizes citric acid as a model system, featuring a complex AB spin system. **Center:** The AI-OC-11.7T pulse sequence is implemented to bridge theory and experiment. **Right:** The waveform of the generated Optimal Control pulse. **B) Comparison between experimental (top) and simulated (bottom) NMR spectra demonstrates the high fidelity of our approach across different quantum states.** the simple in-phase coherence ($I_{1x}$), the anti-phase coherence ($-2I_{1x}I_{2z}$), and the optimized mixed state ($I_{1x} - 2I_{1x}I_{2z}$). **C) Optimization Dynamics:** The fidelity trajectories illustrate smooth convergence towards near-unity performance for all targeted states within 400 epochs. D) **The spectrum evolution in the step-wise spectral improvement**. The signal for the complex state $I_{1x} - 2I_{1x}I_{2z}$ transitions from a near-zero baseline at iteration 10 to a sharp, high-intensity resonance at iteration 300.

**Experimental Validation and Physical Mechanism.** To ensure the strict physical accuracy of our quantum mechanical simulations, the chemical shifts and J-coupling constants for all investigated metabolites (including Citrate, Glutamine, Glutamate, and Cystathionine) were explicitly incorporated into the spin Hamiltonian (Table S1). To validate the efficacy of our framework on physical hardware, we employed citric acid, which is a system characterized by a strongly coupled AB proton network ($H_{AB}$), as a rigorous testbed (Fig. 3A). We implemented the AI-optimized "AI-OC-11.7T" pulse sequences on a high-field spectrometer to bridge the gap between theoretical optimization and experimental realization.

The resulting experimental spectra exhibited remarkable agreement with numerical simulations across a hierarchy of target states (Fig. 3B). While standard in-phase ($I_{1x}$) and anti-phase ($-2I_{1x}I_{2z}$) coherences were generated with high precision, the decisive validation was the successful preparation of a non-intuitive mixed state ($I_{1x} - 2I_{1x}I_{2z}$). This result substantiates our core hypothesis: by directly optimizing the macroscopic spectrum (i.e., maximizing peak height), the algorithm can automatically search within the state space to identify and synthesize the optimal composite state.

As evidenced by the spectral transformation, this specific superposition induces constructive interference, thereby achieving signal maximization in strongly coupled systems. The comparison shows that the signal of the non-intuitive mixed state ($I_{1x} - 2I_{1x}I_{2z}$) is 1.7 times higher than that of the standard in-phase signal ($I_{1x}$).

Furthermore, the optimization dynamics revealed a rapid and smooth convergence, with fidelity trajectories for all target states approaching unity within 400 epochs (Fig. 3C). This stability is visually corroborated by the iterative evolution of the spectral line shape, which transitions from a featureless baseline at epoch 10 to a sharp, high-intensity resonance by epoch 300 (Fig. 3D). Collectively, these results confirm that the spectrum-driven differentiable optimization process allows for the precise inverse design of density matrix trajectories tailored to specific spectral objectives.

Moreover, at a higher field strength of 11.7 T, our framework demonstrated an extraordinary capacity for ultra-precise spectral editing, enabling the selective modulation of individual specific resonances, such as the left-inner or right-outer peaks of citrate, through the deliberate preparation of diverse composite spin states (Fig. S1).

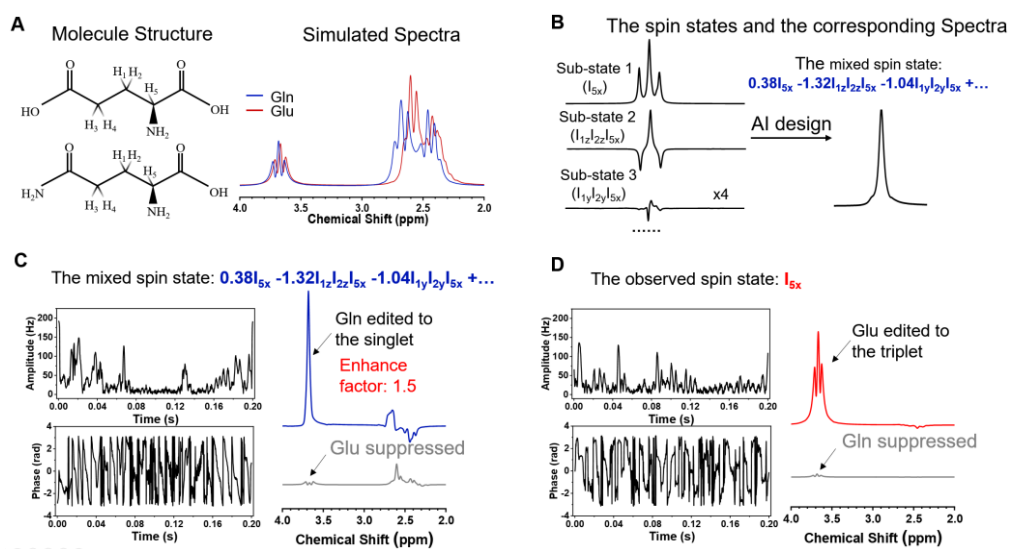

Figure 4. **AI-Enabled Discrimination of Glutamate and Glutamine via Complex Spin-State Engineering.** A) The molecular structures of Glutamate (Glu) and Glutamine (Gln) and the simulated $^1$H spectra. In conventional NMR, their resonances, particularly the alpha-proton signals near 3.7 ppm, exhibit significant overlap, posing a major challenge for precise quantification in vivo. B) Bespoke Spin-State Synthesis. AI has engineered a highly specific mixed spin state (e.g., $0.38I_{5x} - 1.32I_{1z}I_{2z}I_{5x} - 1.04I_{1y}I_{2y}I_{5x} + \ldots$) for Gln. Unlike the characteristic triplet observed for the standard $I_{5x}$ state, this deliberately designed superposition collapses into a sharp singlet, providing a unique and easily identifiable spectral signature. C) Selective Gln Editing. By implementing the AI-optimized pulse sequence (left), Gln is effectively edited into the aforementioned singlet at 3.7 ppm (blue line). Crucially, the same pulse sequence applied to Glu results in near-complete signal suppression (grey line), demonstrating exceptional selectivity even for chemically similar metabolites. D) Selective Glu Editing. Conversely, a separate pulse sequence is designed to target the $I_{5x}$ state of Glu, yielding its characteristic triplet at 3.7 ppm (red line). When this sequence is applied to Gln, the resulting signal is negligible (grey line).

Having established the framework's fidelity on the strongly coupled model system of citric acid (Fig. 3), we next sought to address a critical in vivo challenge: the spectral disentanglement of structurally similar metabolites, Glutamate (Glu) and Glutamine (Gln) (Fig. 4). As illustrated in Fig. 4A, the structural similarity of Glu and Gln results in highly proximal chemical shifts and J-coupling constants, leading to extensive signal overlap in the 3.7 ppm and 2.1 ppm regions11. This makes precise editing of individual components within a crowded spectrum a formidable task for conventional methods.

Addressing this challenge, we employed a spectrum-driven optimization strategy to achieve precise selection and linear modulation of target signals. In the Gln editing task (Fig. 4B, 4C), the algorithm not only effectively extinguished the interfering Glu signal but also modulated the typically complex J-coupled multiplet of Gln into a high-intensity singlet. Quantum mechanical analysis (Fig. 4B) reveals that this phenomenon arises from the algorithm's active introduction of specific high-order coherence terms (e.g., $I_{1z}I_{2z}I_{5x}$, $I_{1y}I_{2y}I_{5x}$, etc.) to form a composite state with the fundamental $I_{5x}$ term. Theoretical analysis further confirms that the AI actively searches for unexploited high-order coherence manifolds. For instance, the system can selectively engineer a targeted spin state (-4 $I_{1z}I_{2z}I_{5x}$) for Glutamine, which macroscopically manifests as a highly distinctive spectral signature with a sharp central peak flanked by two negative lobes (Fig. S7). During evolution, these components induce quantum interference that precisely cancels the intrinsic J-coupling splitting, thereby achieving both spectral simplification and signal enhancement. The comparison shows that the signal of the highly specific mixed spin state (i.e., $0.38I_{5x} - 1.32I_{1z}I_{2z}I_{5x} - 1.04I_{1y}I_{2y}I_{5x} + …$) is 1.5 times higher than that of the standard in-phase signal ($I_{1x}$). It should be noted that the signal enhancement factor of Gln for this pulse sequence does not reach the maximum achievable value (1.8-fold). This is because the sequence was designed to concurrently enhance Gln signals and suppress the Glu signals, which inevitably results in a reduction in the Gln signal enhancement factor.

Furthermore, the framework exhibited exceptional versatility by flexibly switching to the inverse task (Fig. 4D): completely suppressing the Gln signal while preserving the characteristic triplet structure of Glu. These results confirm that the method can keenly capture minute differences in spin Hamiltonians, generating bespoke composite spin states to realize orthogonal selective control over metabolite signals.

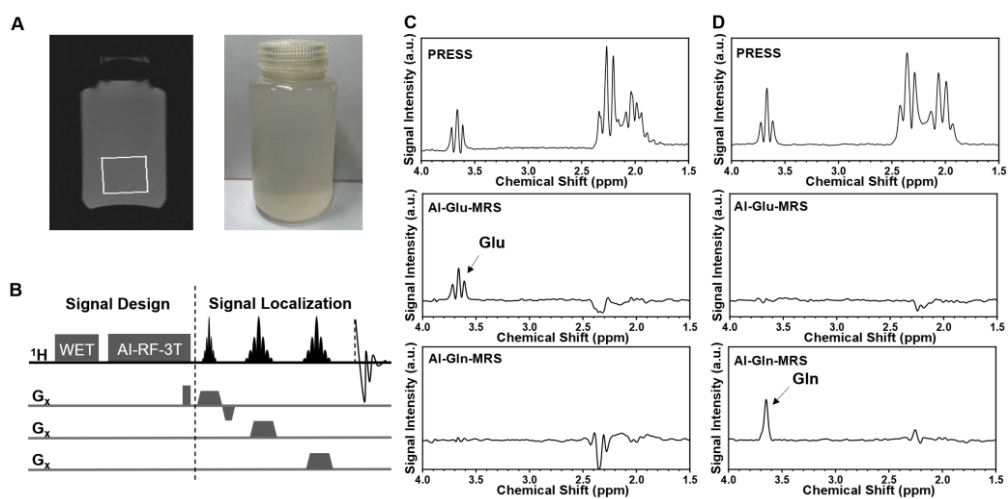

Figure 5. Phantom Validation of the AI-Optimized MRS pulse sequence on a 3T Clinical Scanner. A) The MRI image and photograph of the phantom. B) the framework of the MRS pulse sequence. The module "AI-RF-3T" can be designed to select the signals of different targeted molecules. In this work, two modules were designed. One targets

the $I_{5x}$ triplet state for Glu, and the other targets the bespoke mixed spin state (e.g., $0.38I_{5x} - 1.32I_{1z}I_{2z}I_{5x} - ...$) to collapse the Gln signal into a distinctive singlet. C) The spectra acquired on the Glu phantom by using the standard PRESS sequence (top), the AI-Glu-MRS sequence (middle) and the AI-Gln-MRS sequence (bottom); D) The spectra acquired on the Gln phantom by using the standard PRESS sequence (top), the AI-Glu-MRS sequence (middle) and the AI-Gln-MRS sequence (bottom).

**Clinical Translation and In Vivo Application.** To validate the translational feasibility of our method in a real-world hardware environment, we deployed the AI-generated pulse sequences on a standard 3T clinical MRI scanner, utilizing phantom solutions containing Glutamate and Glutamine (Fig. 5A). Architecturally, the optimized RF pulses were seamlessly integrated as selective excitation modules within a standard point-resolved spectroscopy (PRESS) sequence (Fig. 5B), demonstrating the framework's "drop-in" compatibility with existing clinical protocols.

The experimental outcomes (Fig. 5C, 5D) exhibited exceptional fidelity to the theoretical simulations. In the Gln-targeted experiments (Fig. 5D), the bespoke singlet signature—engineered by the algorithm—was clearly resolved in the physical spectrum, with effective suppression of background interference. Conversely, the Glu-targeted sequence successfully recovered the molecule's characteristic triplet structure (Fig. 5C). This striking concordance between simulation and experiment confirms that the spin evolution trajectories predicted by our differentiable physical model are accurately reproduced in the macroscopic reality of a clinical scanner, establishing the immediate potential of these sequences for medical applications.

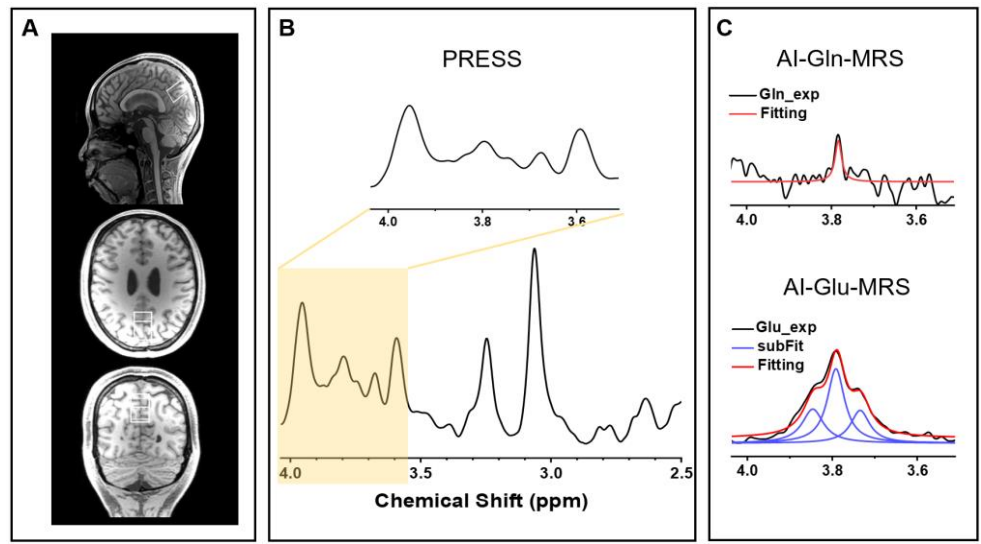

Figure 6. In Vivo Demonstration of Targeted Glu and Gln Discrimination in the Human Brain. A) High-resolution T1-weighted MR image of a healthy subject. The white boxes indicate the Voxel of Interest (VOI) positioned in the parietal lobe. B) The spectrum acquired using the standard PRESS sequence. C) The spectra acquired by using AI-Gln-MRS (Top) and AI-Glu-MRS (Bottom). The experimental data (black) shows an excellent fit (red) with minimal background interference.

The successful phantom experiments (Fig. 5) paved the way for the ultimate test: assessing the framework's robustness and clinical utility in the living human brain (Fig. 6). Compared to the idealized conditions of a phantom, living tissue presents a far more hostile physiological environment, characterized by magnetic susceptibility inhomogeneities and macromolecular baselines, imposing stringent demands on pulse sequence performance. The baseline spectrum acquired with a standard PRESS sequence (Fig. 6B) epitomizes the clinical bottleneck: the severe spectral overlap of Glutamate (Glu) and Glutamine (Gln) in the 3.6-4.0 ppm region creates a broad, featureless envelope that precludes independent quantification.

Against this backdrop of physiological complexity, the AI-generated sequences demonstrated exceptional selectivity and robustness. In the Gln-editing experiment (Fig. 6C, Top), the algorithm successfully "distilled" the Gln signal, normally submerged within the dominant Glu background, into a clean, distinct singlet. Conversely, in the Glu-targeted acquisition (Fig. 6C, Bottom), the sequence effectively suppressed the interfering components while preserving the target's spectral integrity. This ability to maintain high-fidelity separation amidst the complex neurochemical milieu confirms the robustness of our spectrum-driven differentiable physical framework, positioning it as a potent solution for resolving metabolite congestion in clinical MRS.

Beyond the precise spectral disentanglement of overlapping metabolites, we further evaluated the *in vivo* feasibility of our signal enhancement strategy. For Glutamate at 3T, our spectrum-driven optimization achieved a simulated 2.1-fold signal enhancement (Fig. S4), which was subsequently physically validated in phantom experiments, where the Glu triplet was successfully modulated into a sharp singlet with an approximate 1.8-fold intensity gain (Fig. S5). Most importantly, the deployment of this enhancement sequence in the living human brain confirmed that the Glutamate signal could be effectively and robustly enhanced within a real physiological environment, thereby establishing the clinical viability of our goal-driven enhancement approach (Fig. S6).

**Discussion**

Magnetic Resonance Spectroscopy (MRS) has long been regarded as the "gold standard"[20-22] for in vivo metabolic analysis, yet its clinical utility remains strictly constrained by severe spectral congestion and intrinsic low sensitivity. In this study, by introducing a spectrum-driven differentiable physical framework, we demonstrate how to transcend the cognitive boundaries of human intuition and leverage artificial intelligence to identify general solutions within the complex spin dynamics space. By achieving the robust separation of Glutamate (Glu) and Glutamine (Gln) in the human brain at 3T (Fig. 6), we provide evidence that this paradigm not only resolves specific challenges of metabolite overlap but also offers a generalizable pathway for the precise control of high-dimensional quantum systems. The versatility of this goal-driven design philosophy extends well beyond the Glu/Gln system and standard clinical field strengths. As theoretical extensions, our algorithm successfully generated tailored solutions that achieved a three-fold signal enhancement for Cystathionine at 3T (Fig. S2) and a 1.9-fold intensity boost for Glutamine at 5T (Fig. S3), firmly establishing the framework's broad applicability across arbitrary coupled spin topologies and magnetic field environments.

A core finding of this work is the confirmation of the immense, untapped value of "dark" spin states, e.g., the high-order coherence terms long neglected by traditional design, in spectral editing. Conventional pulse sequence engineering is often confined to a "searchlight" heuristic (Fig. 1A), favoring simple single-quantum coherence states (e.g., $I_x$) due to their intuitive physical imagery and ease of manipulation. However, our results indicate that this simplification drastically squanders the exponentially growing information capacity of the state space. Taking the "singletization" of

Glutamine as an example (Fig. 4), the AI did not simply filter the signal but constructed a non-intuitive composite entangled state. Quantum mechanical analysis reveals that the high-order coherence terms actively introduced by the algorithm (e.g., $I_{1z}I_{2z}I_{5x}$) induced precise quantum interference effects during evolution, which destructively interfere with the scalar coupling (J-coupling) splitting. This mechanism is analogous to "dynamical decoupling" in quantum computing, yet its purpose here is to reshape the macroscopic observable spectral lineshape.

Methodologically, the success of this framework stems from transforming pulse design from an ill-posed inverse problem into a well-posed forward optimization problem. In the traditional workflow, researchers must first prescribe a target quantum state and then attempt to deduce RF parameters, often leading to a "theoretical perfection but physical impossibility" dead end. In contrast, our end-to-end differentiable architecture (Fig. 2) establishes a direct gradient pathway from RF parameters to the final spectrum. By embedding the Liouville-von Neumann equation into the computational graph, the algorithm implicitly guarantees the physical realizability of all intermediate states during the search. This "what-you-see-is-what-you-get" closed-loop mode allows us to define loss functions directly based on final clinical requirements (e.g., maximizing resolution, minimizing specific backgrounds) without concerning ourselves with the specific forms of intermediate quantum states, thereby fully unleashing the degrees of freedom in pulse sequence design.

The successful translation from phantom (Fig. 5) to the human brain (Fig. 6) confirms the robustness of this method in real physiological environments. Inevitable magnetic field inhomogeneities ($B_0/B_1$) and complex background signals in living tissue are typically the "killers" of precision pulse sequences[23-25]. However, our AI-optimized sequences exhibited exceptional performance in the human parietal lobe, successfully "extracting" clear metabolite signatures from the originally chaotic 3.7 ppm region. This success is attributed to the robustness constraints introduced during training, which force the AI to seek "broad solutions" that are insensitive to experimental errors. This ability to maintain high fidelity amidst complex interference marks the technology's potential to bridge the "translational gap" and holds promise for directly upgrading existing clinical MRS protocols.

Despite these exciting results, this study is not without limitations. Current optimization targets specific metabolite systems, and computational complexity remains a challenge for large macromolecular systems with high spin counts (>10). Furthermore, while we controlled the specific absorption rate (SAR) via penalty terms, restrictions on RF energy will be more stringent in ultra-high field (e.g., 7T/11.7T)[26, 27] applications. Future work will explore extending this framework to broadband pulse design and multidimensional spectroscopy (2D NMR). Looking further ahead, this "grey-box" strategy, combining First Principles with Deep Learning, is applicable not only to magnetic resonance but also offers a new paradigm for complex system control in fields such as quantum sensing and quantum computing.

**References:**

(1) Oz, G.; Alger, J. R.; Barker, P. B.; Bartha, R.; Bizzi, A.; Boesch, C.; Bolan, P. J.; Brindle, K. M.; Cudalbu, C.; Dinçer, A. Clinical proton MR spectroscopy in central nervous system disorders. *Radiology* **2024**, *270*. DOI: 10.1148/radiol.13130531.

(2) Wilson, M.; Andronesi, O.; Barker, P. B.; Bartha, R.; Bizzi, A.; Bolan, P. J.; Brindle, K. M.; Choi, I. Y.; Cudalbu, C.; Dydak, U.; et al. Methodological consensus on clinical proton MRS of the brain: Review and recommendations. *Magn Reson Med* **2019**, *82* (2), 527-550. DOI: 10.1002/mrm.27742.

(3) Near, J.; Harris, A. D.; Juchem, C.; Kreis, R.; Marjanska, M.; Oz, G.; Slotboom, J.; Wilson, M.; Gasparovic, C. Preprocessing, analysis and quantification in single-voxel magnetic resonance spectroscopy: experts' consensus recommendations. *NMR Biomed* **2021**, *34* (5), e4257. DOI: 10.1002/nbm.4257.

(4) Harris, A. D.; Saleh, M. G.; Edden, R. A. Edited (1) H magnetic resonance spectroscopy in vivo: Methods and metabolites. *Magn Reson Med* **2017**, *77* (4), 1377-1389. DOI: 10.1002/mrm.26619.

(5) Belkić, D.; Belkić, K. Automatic self-correcting in signal processing for magnetic resonance spectroscopy: noise reduction, resolution improvement and splitting overlapped peaks. *Journal of Mathematical Chemistry* **2019**, *57* (9), 2082‐2109. DOI: 10.1007/s10910-019-01060-x.

(6) Saleh, M. G.; Oeltzschner, G.; Chan, K. L.; Puts, N. A. J.; Mikkelsen, M.; Schar, M.; Harris, A. D.; Edden, R. A. E. Simultaneous edited MRS of GABA and glutathione. *Neuroimage* **2016**, *142*, 576-582. DOI: 10.1016/j.neuroimage.2016.07.056.

(7) F, S.; T, N.; U, K.; O, L. Lactate quantification by means of press spectroscopy--influence of refocusing pulses and timing scheme. *Magn Reson Imaging* **1995**, *13* (2), 309-319. DOI: 10.1016/0730-725x(94)00104-b.

(8) Harris, L. M.; Tunariu, N.; Messiou, C.; Hughes, J.; Wallace, T.; DeSouza, N. M.; Leach, M. O.; Payne, G. S. Evaluation of lactate detection using selective multiple quantum coherence in phantoms and brain tumours. *NMR Biomed* **2015**, *28* (3), 338-343. DOI: 10.1002/nbm.3255.

(9) J, P.; P, L. R.; D, N.; A, M. Parameter relations for the Shinnar-Le Roux selective excitation pulse design algorithm (NMR imaging). *IEEE Trans Med Imaging* **1991**, *10* (1), 53-65. DOI: doi: 10.1109/42.75611.

(10) Rund, A.; Aigner, C. S.; Kunisch, K.; Stollberger, R. Magnetic Resonance RF Pulse Design by Optimal Control With Physical Constraints. *IEEE Trans Med Imaging* **2018**, *37* (2), 461-472. DOI: 10.1109/TMI.2017.2758391.

(11) Dhamala, E.; Abdelkefi, I.; Nguyen, M.; Hennessy, T. J.; Nadeau, H.; Near, J. Validation of in vivo MRS measures of metabolite concentrations in the human brain. *NMR Biomed* **2019**, *32* (3), e4058. DOI: 10.1002/nbm.4058 Medline.

(12) Lally, N.; An, L.; Banerjee, D.; Niciu, M. J.; Luckenbaugh, D. A.; Richards, E. M.; Roiser, J. P.; Shen, J.; Zarate, C. A., Jr.; Nugent, A. C. Reliability of 7T (1) H-MRS measured human prefrontal cortex glutamate, glutamine, and glutathione signals using an adapted echo time optimized PRESS sequence: A between- and within-sessions investigation. *J Magn Reson Imaging* **2016**, *43* (1), 88-98. DOI: 10.1002/jmri.24970.

(13) Choi, C.; Coupland, N. J.; Bhardwaj, P. P.; Malykhin, N.; Gheorghiu, D.; Allen, P. S. Measurement of brain glutamate and glutamine by spectrally-selective refocusing at 3 Tesla. *Magn Reson Med* **2006**, *55* (5), 997-1005. DOI: 10.1002/mrm.20875 Medline.

(14) Ramadan, S.; Lin, A.; Stanwell, P. Glutamate and glutamine: a review of in vivo MRS in the human brain. *NMR Biomed* **2013**, *26* (12), 1630-1646. DOI: 10.1002/nbm.3045.

(15) Xin, J. X.; Wei, D. X.; Ren, Y.; Wang, J. L.; Yang, G.; Zhang, H.; Li, J.; Fu, C.; Yao, Y. F. Distinguishing glutamate and glutamine in in vivo (1) H MRS based on nuclear spin singlet order filtering. *Magn Reson Med* **2023**, *89* (5), 1728-1740. DOI: 10.1002/mrm.29562.

(16) Hancu, I.; Port, J. The case of the missing glutamine. *NMR in Biomedicine* **2011**, *24* (5), 529-535. DOI: 10.1002/nbm.1620.

(17) Paszke, A.; Gross, S.; Chintala, S.; Chanan, G.; Yang, E.; DeVito, Z.; Lin, Z.; Desmaison, A.; Antiga, L.; Lerer, A. Automatic differentiation in PyTorch. In NIPS 2017, Long Beach, CA, USA; 2017.

(18) Hodgkinson, P.; Emsley, L. Numerical simulation of solid-state NMR experiments. *Nuclear Magnetic Resonance Spectroscopy* **2000**, *36* (3), 201-239. DOI: 10.1016/S0079-6565(99)00019-9.

(19) Khaneja, N.; Reiss, T.; Kehlet, C.; Schulte-Herbruggen, T.; Glaser, S. J. Optimal control of coupled spin dynamics: design of NMR pulse sequences by gradient ascent algorithms. *J Magn Reson* **2005**, *172* (2), 296-305. DOI: 10.1016/j.jmr.2004.11.004.

(20) Sjobakk, T. E.; Johansen, R.; Bathen, T. F.; Sonnewald, U.; Kvistad, K. A.; Lundgren, S.; Gribbestad, I. S. Metabolic profiling of human brain metastases using in vivo proton MR spectroscopy at 3T. *BMC Cancer* **2007**, *7*, 141. DOI: 10.1186/1471-2407-7-141.

(21) Howe, F. A.; Barton, S. J.; Cudlip, S. A.; Stubbs, M.; Saunders, D. E.; Murphy, M.; Wilkins, P.; Opstad, K. S.; Doyle, V. L.; McLean, M. A.; et al. Metabolic profiles of human brain tumors using quantitative in vivo 1H magnetic resonance spectroscopy. *Magn Reson Med* **2003**, *49* (2), 223-232. DOI: 10.1002/mrm.10367.

(22) Duarte, J. M. N. Challenges of Investigating Compartmentalized Brain Energy Metabolism Using Nuclear Magnetic Resonance Spectroscopy in vivo. *Neurochem Res* **2025**, *50* (1), 73. DOI: 10.1007/s11064-024-04324-4.

(23) Feldman, R. E.; Balchandani, P. A semiadiabatic spectral-spatial spectroscopic imaging (SASSI) sequence for improved high-field MR spectroscopic imaging. *Magn Reson Med* **2016**, *76* (4), 1071-1082. DOI: 10.1002/mrm.26025.

(24) Xin, L.; Tkac, I. A practical guide to in vivo proton magnetic resonance spectroscopy at high magnetic fields. *Anal Biochem* **2017**, *529*, 30-39. DOI: 10.1016/j.ab.2016.10.019.

(25) Oz, G.; Deelchand, D. K.; Wijnen, J. P.; Mlynarik, V.; Xin, L.; Mekle, R.; Noeske, R.; Scheenen, T. W. J.; Tkac, I.; Experts' Working Group on Advanced Single Voxel, H. M. Advanced single voxel (1) H magnetic resonance spectroscopy techniques in humans: Experts' consensus recommendations. *NMR Biomed* **2020**, e4236. DOI: 10.1002/nbm.4236.

(26) Ibrahim, T. S.; Hue, Y. K.; Tang, L. Understanding and manipulating the RF fields at high field MRI. *NMR Biomed* **2009**, *22* (9), 927-936. DOI: 10.1002/nbm.1406.

(27) Fiedler, T. M.; Ladd, M. E.; Clemens, M.; Bitz, A. K. Safety of Subjects During Radiofrequency Exposure in Ultra-High-Field Magnetic Resonance Imaging. *IEEE Letters on Electromagnetic Compatibility Practice and Applications* **2020**, *2* (3), 85-91. DOI: 10.1109/lemcpa.2020.3029747.

**A Differentiable Physical Framework for Goal-Driven Spin-State Engineering in Magnetic Resonance Spectroscopy**

Gaocheng Fu[1, #], Shiji Zhang[1, #], Kai Huang, Xue Yang[1], Huilin Zhang[1], Daxiu Wei[1]*, Ye-Feng Yao[1,2]*

1. Physics Department & Shanghai Key Laboratory of Magnetic Resonance, School of Physics and Electronic Science, East China Normal University, North Zhongshan Road 3663, Shanghai 200062, P. R. China.

2. Institute of Medical Magnetic Resonance and Molecular Imaging Technology, East China Normal University, 200241, Shanghai, P. R. China.

*: The corresponding authors:
Daxiu Wei: dxwei@phy.ecnu.edu.cn
Ye-Feng Yao: yfyao@phy.ecnu.edu.cn
[#]: Gaocheng Fu and Shiji Zhang contributed equally to this work.

**This file includes:**

1. Figures S1 to S7

2. Table. S1

**Figure S1**

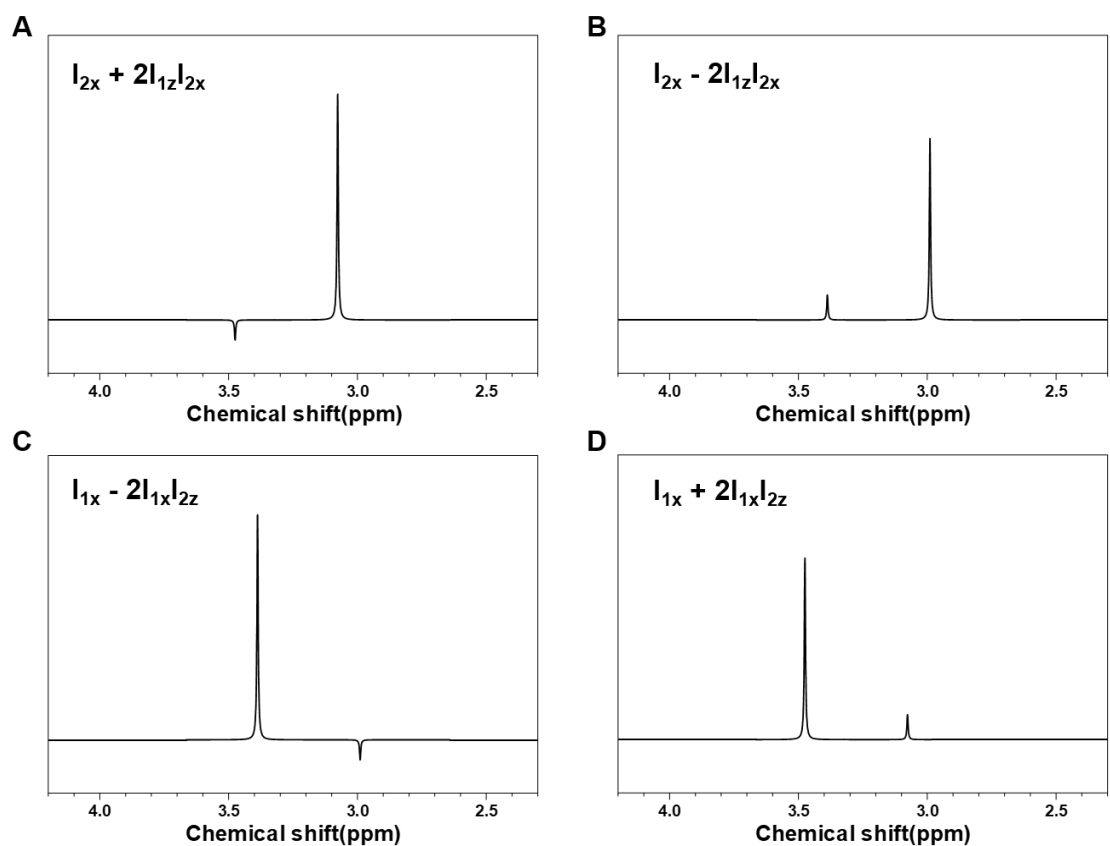

**Figure S1. Simulated lineshape modulation and corresponding spin states of Citrate at 500 MHz.** The subfigures demonstrate the targeted editing of specific resonances: **(A)** the right-outer peak, **(B)** the right-inner peak, **(C)** the left-inner peak, and **(D)** the left-outer peak.

**Figure S2**

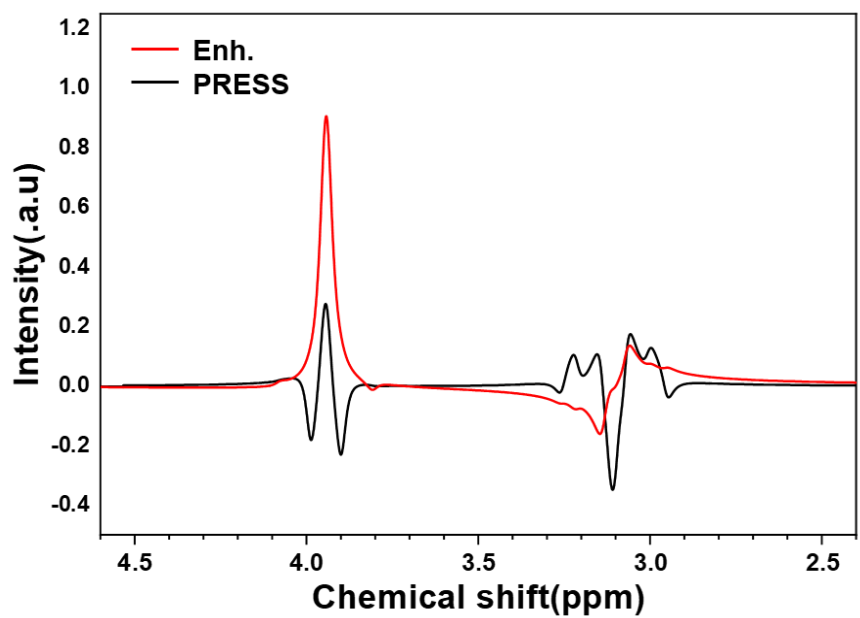

**Figure S2. Simulated signal enhancement of Cystathionine (Cys) at 3T**. Compared to the standard PRESS sequence simulated at an echo time (TE) of 68 ms, the optimized sequence achieves a three-fold increase in signal intensity.

**Figure S3**

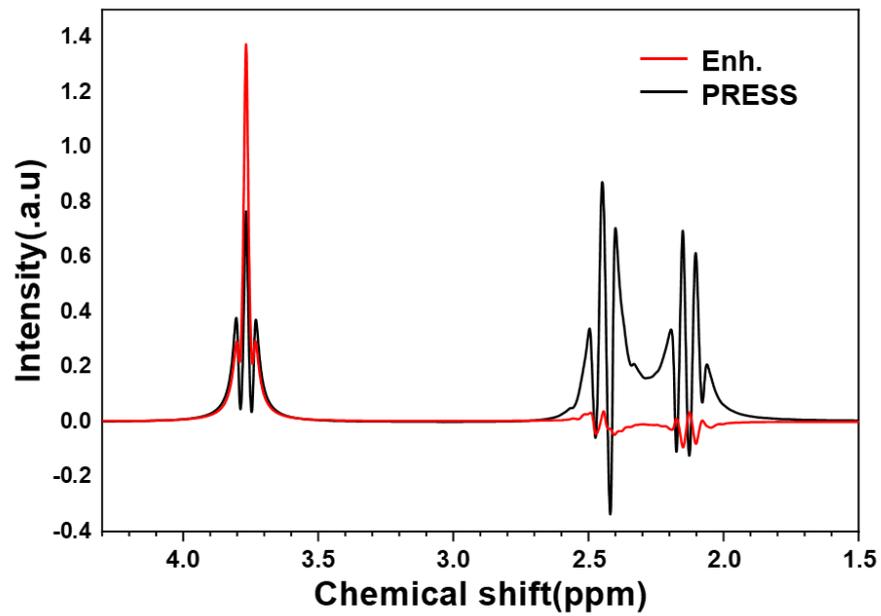

**Figure S3. Simulated signal enhancement of Glutamine (Gln) at 5T.** The AI-optimized sequence yields a 1.9-fold increase in signal amplitude compared to the standard acquisition.

**Figure S4**

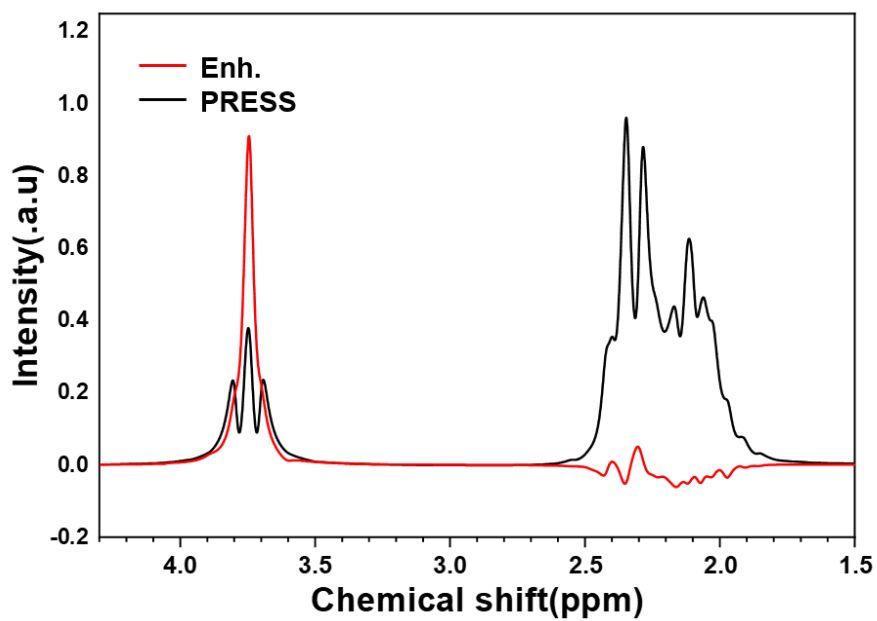

**Figure S4. Simulated signal enhancement of Glutamate (Glu) at 3T.** The optimized sequence demonstrates an approximate 2.1-fold increase in signal amplitude specifically at the 3.7 ppm resonance.

**Figure S5**

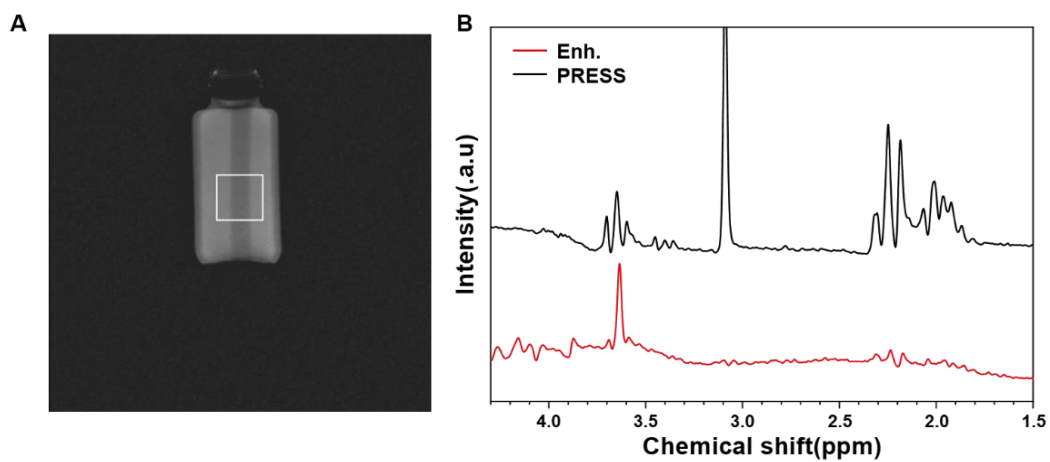

**Figure S5. Phantom validation of Glutamate (Glu) signal enhancement at 3T. (A)** The scout image showing the voxel positioning. **(B)** With the optimization objective set to maximize the signal amplitude at 3.7 ppm, the experimental spectra demonstrate an approximate 1.8-fold enhancement, wherein the characteristic triplet is successfully modulated into a sharp singlet.

**Figure S6**

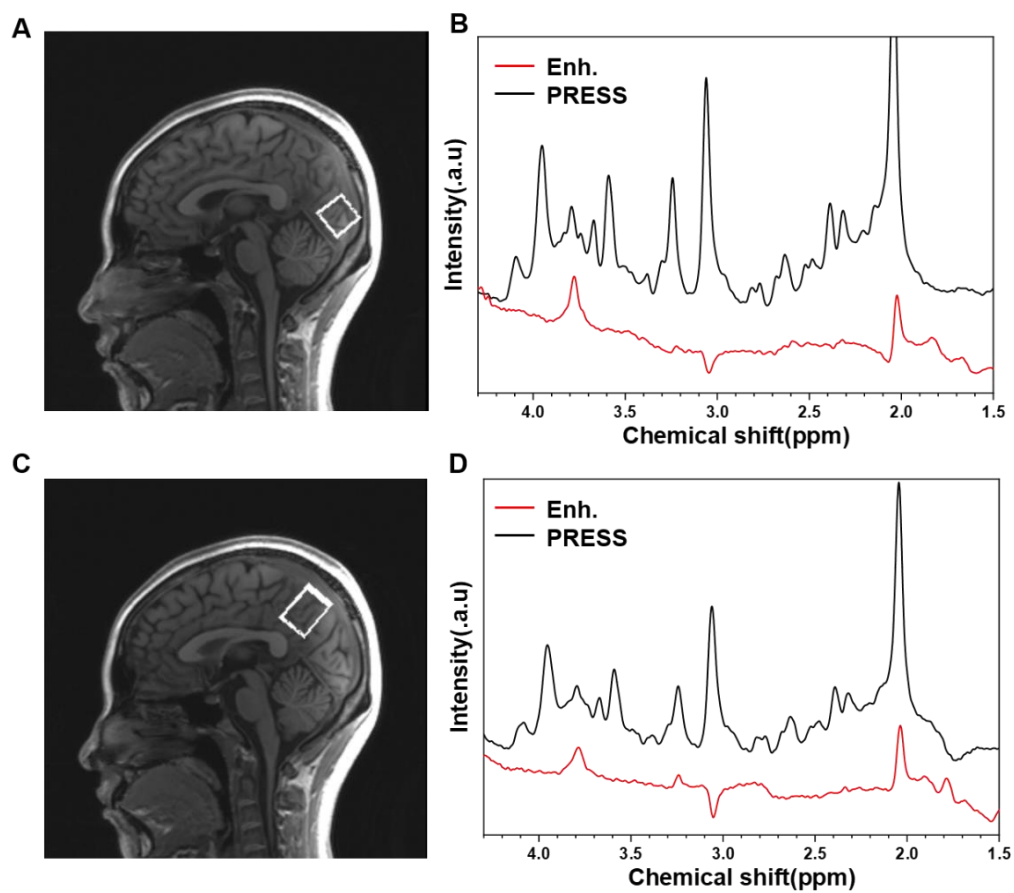

**Figure S6.** *In vivo* **demonstration of the optimized pulse sequence in the human brain.** The scout images and corresponding localized spectra were acquired from the occipital lobe **(A, B)** and the parietal lobe **(C, D)**. The results indicate a consistent and effective signal enhancement of Glutamate across both Volumes of Interest (VOIs).

**Figure S7**

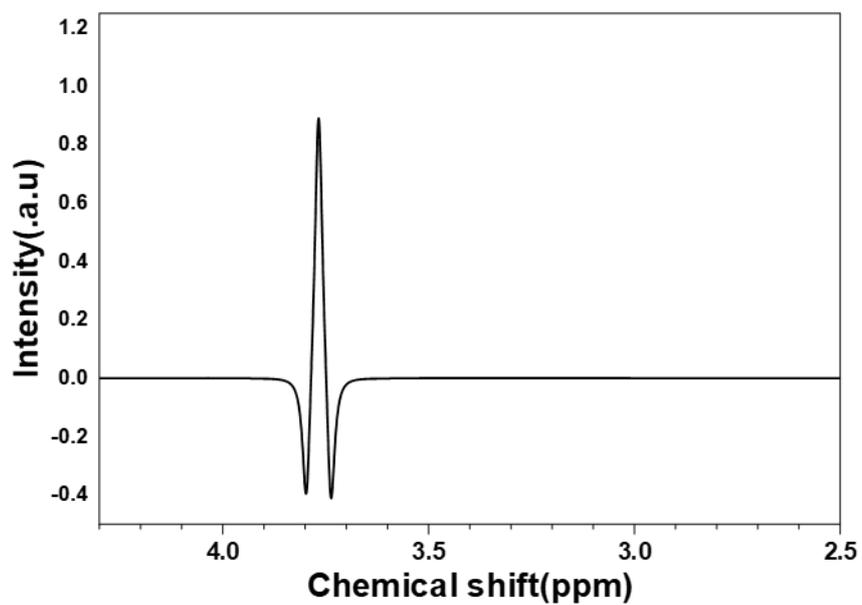

**Figure S7. Spectral signature of a specifically engineered high-order spin state of Glutamine.** The unique lineshape—characterized by a high central peak flanked by two negative peaks—corresponds to the targeted high-order spin state $-4I_{1z}I_{2z}I_{5x}$.

**Table S1. Chemical shifts and J-coupling constants for the simulated metabolites.**

| | |
|---|---|
| **Citric acid** | **Chemical Shift** (ppm): 2.66, 2.53 <br> **J coupling constant** (Hz)**:** J12 = -17.23 |
| **Glutamine** | **Chemical Shift** (ppm): 2.12, 2.13, 2.43, 2.45, 3.76 <br> **J coupling constant** (Hz)**:** J12 = -14.45, J13 = 6.18, J14 = 9.4, J15 = 6.71, J23 = 9.34, J24 = 6.32, J25 = 5.91, J34 = -15.61 |
| **Glutamic** | **Chemical Shift** (ppm): 2.04, 2.12, 2.33, 2.35, 3.74 <br> **J coupling constant** (Hz)**:** J12 = -14.76, J13 = 6.28, J14 = 8.7, J15 = 7.33, J23 = 6.28, J24 = 8.77, J25 = 4.65, J34 = -16.03 |
| **Cystathionine** | **Chemical Shift** (ppm): 3.07, 3.13, 3.94 <br> **J coupling constant** (Hz)**:** J12 = 14.75, J13 = 7.25, J23 = 4.32 |